\documentclass[iop]{emulateapj}
\usepackage{epsfig}
\usepackage{natbib}
\usepackage{graphicx}
\usepackage{xcolor}

\begin{document}

\title[Research Output from Lick Observatory for 1965 - 2019]{Research Output Between 1965 and 2019 from the Use of Telescopes at Lick Observatory}

\author{Graeme H. Smith$^1$ and Matthew Shetrone$^1$}

\affiliation{$^1$ University of California Observatories/Lick Observatory, University of California, Santa Cruz, CA, 95064, USA}
%\address{$^1$ University of California Observatories/Lick Observatory, University of California, Santa Cruz, CA, 95064, USA}
%\ead{graeme@ucolick.org}
%\ead{mshetrone@ucolick.org}
\vspace{10pt}
%\begin{indented}
%\item[]August 2020
%\end{indented}

\begin{abstract}
The productivity of Lick Observatory (LO) is reviewed over the years from 1965 to 2019, a 55 yr period which commences with the Shane 3 m telescope being the second-largest astronomical reflector in the world, but transitions into the
era of 10 m ground-based optical telescopes. The metric of productivity used here is the annual number of refereed papers within which are presented results that are based at least in part on observations made with the telescopes of LO on Mount Hamilton. Criteria are set forth that have guided the counting of this metric. A bibliography of papers pertinent to observations from Lick Observatory has been compiled, and is made available through a NASA ADS library.

The overall productivity of the observatory, counting all telescopes, went through a broad maximum between the years 1975 and 1982. This period also corresponds to a maximum in productivity of the Shane 3 m telescope. An author network shows that this period is attended by the introduction of digital detector systems at LO, particularly at the Shane telescope. Following 1983 the overall productivity of LO shows a net long-term decrease but with two other lesser peaks superimposed on that decrease. A slightly smaller peak occurs around 1996 and is associated with programs taking advantage of CCD spectrometers at both cassegrain and coud\'{e} foci of the Shane telescope. A third lesser peak around 2012 can be attributed to a rise in extragalactic supernova studies originating out of UC Berkeley. Author networks serve to document the UC astronomical communities that were using LO telescopes at these peak times. Institutional affiliations of first authors are documented.
\end{abstract}

%\submitto{\pasp}
%\maketitle

\maketitle

\section{Introduction}

Lick Observatory (LO) was established during the decade of the 1880s. The observatory was made possible by the scientific philanthropy of James Lick, under circumstances that have been well chronicled by \citet{Wright1987} and \citet{Osterbrock1988}. The first telescope to be mounted in a dome atop Mount Hamilton was a 12 in Clark refractor, which was used for observation of a transit of Mercury in 1881 November \citep{Neubauer1950}, a mere two weeks after the Earp brothers and Doc Holliday had faced down the Clanton-McLaury boys at the O.~K. Corral. A 36 in Great Refractor, at the time the largest refracting telescope in the
world, saw first light early in 1888, not quite a decade before the discovery
of the electron was announced by J.~J. Thomson. Lick Observatory has now been
the site of astronomical observations for over 130 yr, with a history that
straddles three different centuries. It is the oldest continuously-staffed
observatory to have been constructed on a relatively high mountain top in the
United States. Research programs are still being actively supported by
telescopes at Lick Observatory, although the suite of telescopes on Mount
Hamilton has changed over the decades. Such considerations make Lick
Observatory an interesting case study for tracing the long-term research
output of an astronomical facility.

A variety of sources have chronicled many of the people and events associated
with the foundation and history of Lick Observatory \citep{Neubauer1950,
Wright1987, Osterbrock1988}. Summaries of research activities during
the first fifty years of the observatory were written by \citet{Aitken1928} and \citet{Moore1938}. By comparison, in this paper the research output from Lick Observatory telescopes is traced over a more recent 55 yr period. Our survey covers the years from 1965 to 2019 inclusive. Prior to this time some research programs of Lick astronomers were being reported in several series of Observatory-sponsored publications, notably the Lick Observatory Bulletins (of which twenty volumes were published between 1901 and 1951), and the Publications of Lick Observatory, without necessarily being accompanied by duplicate publication in a refereed journal. Thus we have restricted the start date of our survey period to avoid the era in which a consideration of papers published in refereed journals only would omit some significant works.

The scientific output of the Shane 3 m telescope on Mount Hamilton has
previously been studied for the years 1980-1981, 1990-1991, and 2000-2001 by
\citet{Abt1985}, \citet{Trimble1995}, and \cite{Trimble2005} respectively, as 
part of surveys of the productivity of large telescopes both globally and in the United States. By contrast, the goal of the present paper is to focus on Lick Observatory itself, and to provide a time-resolved panorama of the published output from the observatory over a period encompassing more than half a century.

\section{Setting the Scene}

In a sense the Lick Observatory represents an ecosystem in which different
telescopes have been tailored toward different types of research programs.
At the beginning of the 55 year survey period the Shane 3 m telescope had been
acquiring observations for just on six years, having been commissioned in 1959. Thus the earliest years of our survey period date to an era in which the Shane telescope was the forefront research instrument of astronomers within the University of California (UC) system, providing imaging, spectroscopic and
photometric observations. However, it was by no means the only telescope on
Mount Hamilton that was supporting active research programs. Astrometry via
both photographic and visual techniques was being pursued with the 36 in Great
Refractor, which also had a considerable history of spectroscopy. A 20 in
Double Astrograph was engaged in a large photographic survey of stellar proper
motions. Photometry via either photographic or photoelectric devices was being
carried out at the 0.9 m Crossley reflector and a 0.6 m Boller and Chivens
reflector. The Clark 12 in refractor was also still in use for photoelectric
and photographic observations. However, in 1979 this telescope was replaced
by installation of the 1 m Anna Nickel reflector within the Clark dome.
A listing of telescopes that have been in use on Mount Hamilton during the 
1965-2019 period of our literature survey is given in Table \ref{table1}.

Moving through the 55 year survey period a number of factors came into play
that caused UC astronomers to increasingly diversify their observational
activities beyond the resources on Mount Hamilton. As the skies of the
San Jose area became brighter, and national facilities with 4 m telescopes
were developed at the darker sites of Kitt Peak National Observatory and
Cerro Tololo Inter-American Observatory, the role of the Shane 3 m telescope
began to be impinged upon \citep[e.g.,][]{Trimble2005}. The telescope was
kept vital by the development of state-of-the-art spectrometers combined with
electronic detectors \citep[e.g.,][]{Wampler1966, Robinson1972, Miller1979, 
Abt1985, Vogt1981, Vogt1987}, enabling it to make advances in high-resolution
coud\'{e} spectroscopy and both sky-subtracted and flux-calibrated cassegrain
spectroscopy at lower resolution.

The initial suite of instruments used with the Shane telescope were a coud\'{e} spectrometer and various photographic and photoelectric (pe) devices placed at prime focus. The cassegrain focus of the Shane telescope was not modified for spectroscopy until 1969. A summary of the main instruments that have been used on the 3 m telescope over the 1965-2019 period is given in Table \ref{table2}. The third column lists the main years in which data from each instrument appeared in refereed publications. The image-tube, image-dissector (ITS) scanner was first installed with a cassegrain spectrograph on the Shane telescope in 1970. However, by 1975 that spectrograph had been replaced by another with a more suitable optical design employing a Schmidt camera \citep{Miller1980}. A spectropolarimeter was added along with the new spectrograph, and thus commenced a multi-decade capability of spectropolarimetry that continues on the Shane telescope to this day. The ITS system was eventually replaced by CCD detectors and re-designed spectrographs, with the current cassegrain system being the Kast double beam spectrograph. The Kast spectrograph was optimized for long-slit throughput across the visible spectrum down to the atmospheric cutoff at 3000 \AA, and it has a suite of grisms and gratings giving a range of resolutions. The original coud\'{e} optical train is now rarely used, having been replaced by echelle systems, the latest of which is the Hamilton echelle spectrometer.

The 1990s saw two major developments that redirected UC astronomers well beyond the skies of Mount Hamilton, the Hubble Space Telescope was launched in 1990, while the Keck 1 and 2 telescopes were commissioned on Maunakea in 1993 and 1996 respectively.
Lick Observatory responded to these developments in part by moving into new
specialized directions. In the 1990s and the first decade of the twenty-first
century the Shane telescope became a test-bed for the development of a natural
and laser guide star adaptive optics instrument \citep[e.g.,][]{Max1997}.
Infrared cameras and spectrographs compatible with the Shane telescope were
developed at the UCLA, UCB, UCSC, and UCSD campuses \citep[e.g.,][]{Rudy1991,Gilmore1994,McLean1994,McLean1995,Lloyd2000}. Some 
of these infrared devices would go on to become general user instruments available to all UC astronomers, such as the UCLA Gemini camera, whereas others remained the
province of the groups that built them. Two new special-purpose telescopes
were established on Mount Hamilton in the 21st century, namely the 0.76 m
Katzman Automatic Imaging Telescope (KAIT) \citep{Filippenko2001} and the 2.4 m Automated Planet Finder (APF) telescope \citep{Vogt2014}.   It is of note that these last two special-purpose telescopes have a different operational model to the classically scheduled Shane and Nickel telescopes; both are robotic queue-scheduled telescopes.

\section{Description of the Literature Survey}

A search for papers in which are reported observations and/or data acquired
from telescopes at Lick Observatory was made for the period covering 1965-2019
by using the SAO/NASA Astrophysics Database System \citep[ADS;][]{Kurtz2000}.
Papers counted for this survey have been published in refereed journals only.
Observatory reports or technical manuals written, for example, for internal
distribution within the UC system have been not counted. Papers published in
edited conference proceedings have not been counted unless those proceedings
appeared in a refereed journal. In a few exceptional cases, such as
\cite{Shane1967}, for example, important works that did not appear
originally in refereed literature but in the series of Publications of
Lick Observatory have been included in our survey.

\subsection{Criteria for the Survey}

A compilation of over 3200 papers was produced as a result of the current
survey. Some basic criteria were used for deciding upon inclusion of a paper
within this compilation. Papers that report results from new observational
data form the main stay of our survey. No regard was given to whether a paper
is based entirely on data derived from Lick Observatory, or whether it has
incorporated otherwise-unpublished Lick data into a larger observational
program that utilized many different telescopes. If a refereed paper includes
at least some new data from observations made at Lick Observatory it has
generally been counted.

In cases where a series of clearly connected papers has been published by a
group of authors based on a large research program, with the first paper in
the series being predominantly a report on the acquisition and reduction of
the acquired data, while scientific analyses of derived measurements are
described in following papers in the series, the multiple papers from such a
series were generally counted. In a somewhat different context, papers have
been counted if they report results derived from a reanalysis or re-reduction
of Lick-derived data that were initially described in a refereed paper by
unrelated authors. By contrast, we have tended not to count papers when
authors have drawn directly from tables of measurements within previous
Lick-based papers, although an exception has sometimes been made when there
is evidence of a collaboration between authors of the two papers.

Despite the sincere intentions of the authors the current literature survey does not claim to be complete. Even with the use of the ADS some relevant papers may have been overlooked. Furthermore, although the above selection criteria are formalized in principle, it has been our experience that in practice their consistent application can be subjective. However, an effort has been made to adopt a set of criteria that avoid any need to make patently subjective judgments, such as whether the presence of LO data in a paper made a significant or meaningful contribution to the conclusions or analysis presented in that paper.

\subsection{An ADS Library of ``Lick Observatory'' Related Papers}

A database of Lick-related refereed papers compiled from our literature survey has been made available through the ADS Public Library system 
(https://ui.adsabs.harvard.edu/public-libraries/GiKGUgHbTHu4d6UfaSYNog)
and can be accessed in an ADS search by using the ADS search 
"docs(library/GiKGUgHbTHu4d6UfaSYNog)"
This can be combined with normal ADS search parameters, e.g. 
"docs(library/GiKGUgHbTHu4d6UfaSYNog) AND author:"Vogt" AND year:2013-2015".
In that example there would be two non-refereed papers and six refereed papers
returned; if only refereed papers were desired the search could be 
narrowed down to have that property by adding "+property:refereed" to 
the search.    

\subsection{Some Limitations of the Survey}

By restricting our survey to papers published in refereed journals there are some creative areas of research associated with Lick Observatory that will be underrepresented. One notable example consists of instrumentation developments that have been reported in conference proceedings, particularly those published by the International Society for Optics and Photonics (SPIE). In order to provide a fuller picture of the active instrumentation research and development programs that have contributed to Lick Observatory, a partial list of SPIE papers has been added to the on-line ADS library referenced above. This literature provides a record of instrument building throughout the University of California Observatories system, including UCSC, UC Berkeley, UC Los Angeles, the Lawrence Livermore National Laboratory, and UC San Diego.
\footnote{For example, the development of adaptive optics systems used at Mount Hamilton can be traced through this record (e.g., \cite{Olivier1994}, \cite{Olivier1999}, \cite{Max1997}, \cite{Gavel2003}, \cite{Gavel2016}, \cite{Kupke2012}, and \cite{McGurk2014}).} 
These SPIE papers are not included in the list of refereed publications analyzed in Sections 4-7. 

A survey that is restricted to refereed papers also has other limitations. The impact of projects such as the Shane-Wirtanen galaxy counts \citep[e.g.,][]{Peebles1974,Groth1977,Geller1984} will be underrepresented because only the original data papers are included according to our survey criteria. The legacy of a program such as the Lick Observatory Supernova Search \citep{Filippenko2001,Leaman2011,Li2011}, which has produced numerous IAU Circulars and other discovery announcements, is not fully encompassed by the refereed literature. As another example, use of the 36 in Great Refractor in the 1960s by geologists of the United States Geological Survey contributed to a variety of products not found in refereed literature, such as selenographic maps \citep{Wilhelms1993}. 
Lick Observatory reports such as those once published by the American
Astronomical Society \citep[e.g.,][]{Whitford1969, Osterbrock1982} have not been included in the current survey.

\section{Basic Trends Within the Survey}

\begin{figure}[ht]
\centering
\includegraphics[width=.50\textwidth]{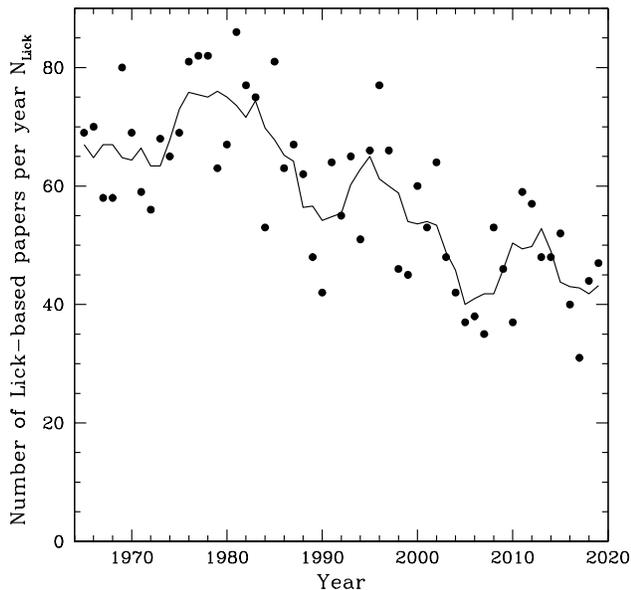}
\caption{The number of papers $N_{\rm Lick}$ published per year within 
which data obtained from telescopes of Lick Observatory were reported during 
the period 1965-2019. Boxcar averages of the annual number of papers were calculated employing a bin of 5 yr. These average points are not plotted individually but are connected in the figure by a solid line, which is intended to be helpful in
identifying long-term trends. }
\label{fg1}
\end{figure}

The compilation of papers that resulted from our survey has been used to count
the number of refereed papers per year in which are reported at least some
observational material derived from telescopes at Lick Observatory. This annual number, denoted in the following text as $N_{\rm Lick}$, is plotted for the period from 1965 to 2019 in Figure \ref{fg1}. The highest number of Lick-based papers was attained in 1981, within a period of particularly high productivity extending from
1975 to 1986 during which there are a number of years when the annual output 
equalled or exceeded 80 papers (the highest levels achieved in our survey period). 
This maximum near 1981 is referred to below for convenience as the ``first peak'' in the 1965-2019 publication record of Lick Observatory.\footnote{We thank David Soderblom for originally drawing our attention to this period in a conversation with GHS a number of years ago.} After this period of high output an upper envelope to the behavior of $N_{\rm Lick}$ in Figure \ref{fg1} shows a decline since 1986. Around 2011 there was a short interval in which the annual number of papers were comparable to some years between 1965 and 1972. However, since 2012 the annual output $N_{\rm Lick}$ has not reached the levels of the 1965-1975 period. There can also be identified a lower envelope to the data in Figure \ref{fg1}, corresponding to years in which the annual number of papers was at a relative minimum. This minimum envelope extends from $\sim 58$ papers per year around 1970 to 31 papers in 2017. Within the bounds of the upper and lower envelopes there is marked year-to-year scatter. The counts for $N_{\rm Lick}$ plotted in Figure \ref{fg1} thus evince both substantial short-term scatter and more gradual long-term trends. The average number of papers per year that are based at least in part on data from Lick Observatory telescopes during the period 1965-2019 is 59 with a standard deviation of 14.

Differences in $N_{\rm Lick}$ of up to 20 or more papers have occurred between consecutive years, and such fluctuations are distributed throughout the 55 yr survey period. The average rate of change has been $-0.4$ papers per year, with a standard deviation of 12.1, for the period 1966-2019.

In order to search for long-term trends in the paper counts the data from Figure \ref{fg1} were subjected to a boxcar smoothing with a bin of 5 yr. The resulting smoothed averages are connected by a solid line in Figure \ref{fg1}. There is an overall decline in the number of papers per year since about 1980, consistent with the negative average rate of change in $N_{\rm Lick}$ noted above. However, on top of this mean trend there is superimposed an almost pseudo-oscillatory variation that is driven by three eras in which the smoothed number of papers per year went through a local maximum. Over the period 1965-1970 the annual total of papers was roughly constant. Thereafter is evident an upturn in the yearly number of papers from 1970 to 1975, followed by a period from 1975 to around 1983 in which the annual number of papers reached the highest levels within the entire 55 yr survey being considered here. The yearly paper count thereafter decreased to a local minimum near 1990, after which there followed a rise to a local maximum at around 1996, although this second peak did not attain the highest levels of the first one. A third, yet even lower, local maximum occurred around 2012.  The first and third peaks identified in Figure \ref{fg1} do not seem to be products of an isolated high-productivity year skewing the smoothed data, although the second peak is clearly influenced by a relatively high paper count in 1996.

\begin{figure}[ht]
\centering
\includegraphics[width=.50\textwidth]{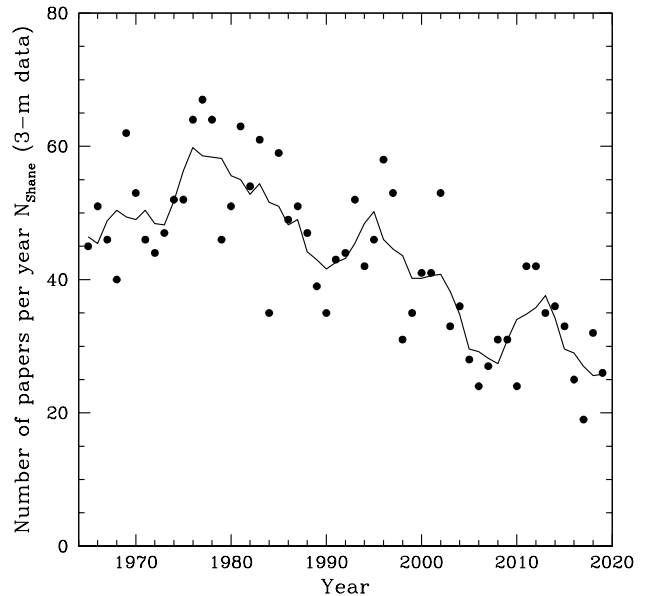}
\caption{Number of papers published per year $N_{\rm Shane}$ that contain results based on observational data acquired with the Shane 3 m telescope for the period 1965-2019. Boxcar averages of the paper numbers were calculated employing a bin of 5 yr, and are connected in the figure by a solid line. The trends seen among the Shane-based papers are very similar to those seen in Figure \ref{fg1} among all Lick-based papers.}
\label{fg2}
\end{figure}

Figure \ref{fg2} shows the annual number of papers since 1965 in which are reported data from the Shane telescope ($N_{\rm Shane}$).  A smoothed version of these counts, again based on a boxcar average over 5 yr intervals, is plotted as a solid line. By and large there are patterns seen in Figure \ref{fg2} that mimic those in Figure \ref{fg1}. The productivity of the Shane telescope has previously been studied for the years 1980-1981 and 1990-1991 by \cite{Abt1985} and \cite{Trimble1995}, respectively, who looked at outputs from a number of large optical telescopes. Figure \ref{fg2} shows an oscillating pattern superimposed on a general decrease since the mid-1980s. A decline in the productivity of the Shane 3 m telescope had previously been identified and commented upon by \cite{Trimble2005} from a comparison of publications records for 1991 and 2001. Overall Figure \ref{fg2} replicates the pseudo-oscillatory pattern seen in Figure \ref{fg1} for Lick Observatory as a whole. Peaks in the annual total paper counts are closely matched by peaks in the output from the Shane telescope.

\begin{figure}[ht]
\centering
\includegraphics[width=.50\textwidth]{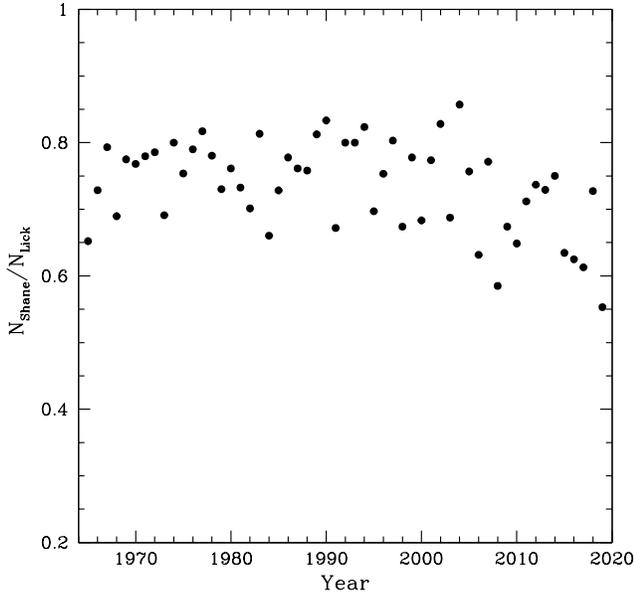}
\caption{Ratio $(N_{\rm Shane}/N_{\rm Lick})$ between the number of
Shane-data papers to all Lick-data papers calculated annually throughout the period 1965-2019. }
\label{fg3}
\end{figure}

A ratio between the number of Lick-wide papers and Shane-based papers per
year, $N_{\rm Shane}/N_{\rm Lick}$, is shown in Figure \ref{fg3}. Between 1965 and
2005 this ratio remained roughly constant on average and fluctuated between
about 0.65 and 0.85. Since 2005 the ratio may show a slight net drop that is discussed in Section 5.4. The suite of active ``smaller'' telescopes on Mount Hamilton has evolved notably in composition during the 55 yr survey period reviewed here. However, collectively this pool of telescopes has maintained a fairly consistent contribution to the published output from Lick Observatory. 

\section{Discussion of the Refereed Publication Record}

\subsection{The Maximum in the 1965-2019 Publication Record}

The annual number of Lick-based papers exceeded 80 in the years 1969, 1976, 1977, 1978, 1981, and 1985, while no year attained 90 papers. The number of papers containing Shane 3 m data in these years is 62, 64, 67, 64, 63 and 59 respectively, corresponding to the highest outputs from the Shane telescope also. The only other year with more than 60 Shane-derived papers is 1983 in which $N_{\rm Lick} = 75$ and $N_{\rm Shane} = 61$. The period from 1973 through 1986 will be referred to herein as the ``first peak'' in the Lick Observatory productivity curve, it is a time of maximal output for Lick Observatory within the past 55 yr, and also coincides with the most productive years of the Shane telescope.

The start of this first-peak period in the LO publication record coincides closely with the introduction of a cassegrain instrument stage for the Shane 3 m telescope \citep{Faulkner1971},  and the development of a multichannel scanner spectrometer, also referred to in the literature as the image-dissector scanner or image-tube scanner (ITS) \citep{Faulkner1971,Robinson1972,Kraft1972,Faulkner1973}, for use at the cassegrain focus. The new facility instrument could supply linear, sky-subtracted, and flux-calibrated spectra of faint galaxies and quasars with the ease of cassegrain access. In the post-1979 years of the first-peak period CCD detectors were introduced to Lick Observatory \citep{Lauer1984}, while a high-resolution cross-dispersed echelle was constructed at the coud\'{e} focus of the 3 m telescope \citep{Soderblom1978} that could be used with either an image-tube image-dissector scanner or a Reticon photodiode array detector \citep{Vogt1981}. 

\begin{figure}[ht]
\centering
\includegraphics[width=.50\textwidth]{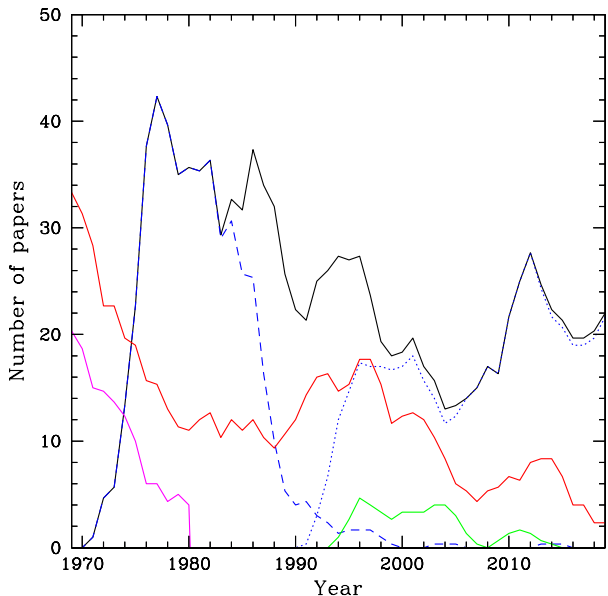}
\caption{Productivity from various instruments on the Shane telescope. Yearly paper numbers have had a boxcar filter of width three years applied in order to partially smooth out the year-to-year fluctuations and thus make long-term trends more apparent. Line type and color denote the instruments in use as follows: solid black (all cassegrain spectrographs), blue dotted (Kast spectrograph), blue dashed (ITS spectrograph), red (coud\'{e} spectrometers), magenta (prime focus), green (UCLA Gemini camera).}
\label{fg4}
\end{figure}

We have attempted to investigate the nature of the first peak in Figures \ref{fg1} and \ref{fg2} by looking at the usage of various instruments on the Shane telescope between 1969 and 2019. Some results are shown in Figure \ref{fg4}, in which is plotted the annual number of papers that incorporate observations from different instrument combinations. The plotted curves show counts that have been smoothed with a running boxcar of bin width equal to 3 yr. These instrument-usage statistics are based on descriptions of observational setups given in published papers, and the degree of detail varies considerably, such that in some cases the data may be lower limits.

A black curve in Figure \ref{fg4} represents the number of papers per year that report spectra obtained at the cassegrain focus of the Shane telescope, whereas
the blue dashed curve represents data from the ITS cassegrain spectrograph and the blue dotted curve the Kast spectrograph. Coud\'{e} spectroscopy at the Shane is depicted by the red curve in Figure \ref{fg4}. At the start of our survey period in 1965 photographic coud\'{e} spectroscopy dominated the output from the Shane telescope, however it declined progressively until circa 1980. Following the introduction of an echelle spectrometer the paper output from the coud\'{e} focus increased into the era of the discovery of exoplanets. The decline in 3 m coud\'{e} use in recent years has accompanied the introduction of the APF telescope. Papers that report data obtained from the use of prime focus instruments are shown as a magenta curve in Figure \ref{fg4}, while the green curve corresponds to observations made with the UCLA Gemini infrared camera.

Figures \ref{fg4} and \ref{fg2} show that the first peak in the productivity curve of Lick Observatory (Figure \ref{fg1}) is clearly associated with a considerable expansion in cassegrain spectroscopic capabilities at the Shane telescope. It is evident from Figure \ref{fg4} that the circa 1973 start of this first peak corresponds to a marked expansion in usage of the ITS on the Shane telescope, particularly upon the introduction of an optimized cassegrain spectrograph in 1975. By the end of the first LO peak period in 1986, data from the ITS and a subsequent 3 m CCD cassegrain spectograph were contributing comparably in the refereed literature. The first peak in the Shane productivity curve (Figure \ref{fg2}) is clearly a cassegrain phenomenon, since papers based on data from both the coud\'{e} spectrometer and the various prime focus instruments exhibit a progressive decrease in number through the years from 1973 to 1986. The new capabilities of an echelle spectrometer introduced during the first peak period do not translate into a maximum in the literature until the era of the second peak in LO productivity (red curve in \ref{fg4}).

Several smaller telescope developments occurred within the era of the first peak.
The Nickel reflector was built and commissioned. In addition, a 0.6 m Coud\'{e} Auxiliary Telescope was placed into service in 1971-1972 that could be used to independently feed the 3 m coud\'{e} system \citep{Whitford1969,Kraft1972}, thereby permitting high-resolution spectroscopy to be carried on through both dark and bright time. Figure \ref{fg5} shows the productivities of the smaller telescopes on Mount Hamilton, which are discussed in more detail in Section 5.4. Suffice it to note here that the combined paper numbers from these telescopes go through a local maximum centered around 1982, partly as a result of output from the Crossley reflector and the introduction of the Nickel telescope. Thus the 1973-1986 ``first peak'' seen in Figure \ref{fg1} really is a maximal time for Lick Observatory across the entire telescope ecosystem.

Instrumentation alone may not be the sole driver behind the relatively high output of the 1969-1985 period. This era commenced about four years after Lick research astronomers ceased being based on-site at Mount Hamilton and relocated to the new Santa Cruz campus of the University of California. This lead to the establishment of a Board of Studies in Astronomy and Astrophysics at UCSC and the start of a new astronomy graduate program that had access to the Mount Hamilton telescopes. It is beyond the scope of this paper to study the evolution of the faculty, postdoctoral, and graduate student populations in the astronomy programs throughout the UC system. However, such a study may provide insights into the behavior of the LO productivity curve.

\subsection{The Second Peak in the Publication Record}

A secondary maximum in the number of Lick and Shane-based papers is seen centered around the years 1995-1996 (Figures \ref{fg1} and \ref{fg2}). A large fraction of Shane papers published around the time of the 1990s secondary maximum are based on two types of observational material, low-resolution spectroscopy made with the Kast double-beam cassegrain spectrograph and the UV Schmidt spectrograph that preceded it \citep[e.g.,][]{Stone1993,Goodrich1994},  and high-dispersion spectra obtained with the Hamilton Echelle Spectrometer, installed in the Shane coud\'{e} focus room \citep{Vogt1987}. In Figure \ref{fg4} local maxima can be seen around 1994-1997 in the number of papers that employ data from either the cassegrain spectrographs or the coud\'{e} spectrometers. In the former case the cassegrain peak coincides with the initial few years of Kast usage. A 1990s local maximum in coud\'{e} output is the result of science programs using the Hamilton echelle spectrometer (Table \ref{table2}). Furthermore the first science papers from the UCLA Gemini infrared camera enter the literature (Figure \ref{fg4}) at the time of the second peak in Shane productivity. There is no obvious maximum in science output from the smaller telescopes on Mount Hamilton at this time (Figure \ref{fg5}). Thus it seems that the second LO productivity peak (Figure \ref{fg1}) marks a coming to fruition of several instruments on the Shane telescope.

Interestingly, the second LO peak occurred around the time that the Keck I telescope went into regular observational use and began to strongly influence the ground-based programs of UC astronomers. The second peak may have been abridged as a consequence of this coincidence. Factors associated with the origin of the second maximum are further addressed in Section 6. 

\subsection{The Third Peak in the Publication Record}

The third peak in the LO and Shane telescope productivity curves around 2012 can also
be viewed in light of the instrument usage shown in Figure \ref{fg4}. Here a local
maximum in Kast cassegrain spectroscopy can be identified, and there is a
modest upturn in science output from the Hamilton echelle as well. In the case of
the third peak it is possible to identify a specific science program that also
shows a marked increase in output around this time. That program is 
discussed in Section 7.2.

\subsection{Contributions of the Smaller Telescopes}

Although the Shane telescope has been the mainstay of observational work at
Lick Observatory it has been complemented by a suite of smaller telescopes. As
has been noted above, the ratio of Shane-facilitated papers $N_{\rm Shane}$ to total LO-papers $N_{\rm Lick}$ held approximately constant over the period from 1965 to 2005, with a possible small average decrease since 2005 (Figure \ref{fg3}). The lack of systematic long-term change in this ratio may reflect the fact that the number of active telescopes on Mount Hamilton held roughly constant throughout the 55 yr survey period. During this period the 12 in Clarke refractor, the Tauchmann 22 in telescope, the twin Astrograph, and the 0.9 m Crossley reflector were effectively retired. The 36 in Great Refractor has ceased to see active research use while it still contributes substantially to public outreach events and activities. The Boller
and Chivens 0.6 m originally on Mount Hamilton has been moved to the Steward
Observatory station on Kitt Peak and has been functioning since 2000 as the
robotic Super-LOTIS (Livermore Optical Transient Imaging System) telescope
\citep{Williams2008}. Research from the Super-LOTIS project has not been
included in the papers counted for this survey.

Three telescopes were added into the Lick Observatory ecosystem over the
survey period, the 1.0 m Anna Nickel reflector, the robotic 0.76 m Katzman
Automatic Imaging Telescope and the 2.4 m Automated Planet Finder
telescope. The Nickel telescope was commissioned in 1979 for general purpose
photometry. The main user instrument is currently a direct-imaging CCD camera. It has
also seen use as a testbed for a variety of instrumentation such as visible
light adaptive optics \citep{Gavel2008}, and optical and near-infrared SETI
\citep{Wright2001,Maire2016}. It was the first telescope on Mount
Hamilton to be enabled for remote observing \citep{Grigsby2008}. The KAIT is
a special-purpose imaging telescope run by the Filippenko group at UC Berkeley
\citep{Filippenko2001}, and is used for the discovery and photometric
characterization of extragalactic supernovae (SNe). It has prompted a considerable
amount of follow-up spectroscopy with the Shane 3 m \citep[e.g.,][]{Silverman2012}.  
Exoplanet science was the driving force behind the installation of
the Automated Planet Finder telescope \citep{Radovan2014,Vogt2014}.
To date it has one Nasmyth instrument, the high-resolution Levy echelle
spectrometer \citep{Vogt2014}. One of a network of remote stations for
monitoring meteor showers (known as CAMS) has also been installed on Mount
Hamilton \citep{Jenniskens2011}. On balance Figure \ref{fg3} indicates that these
newer telescopes on Mount Hamilton have performed roles that have taken over
from those once held by the smaller telescopes that have gone out of service.

\begin{figure}[ht]
\centering
\includegraphics[width=.50\textwidth]{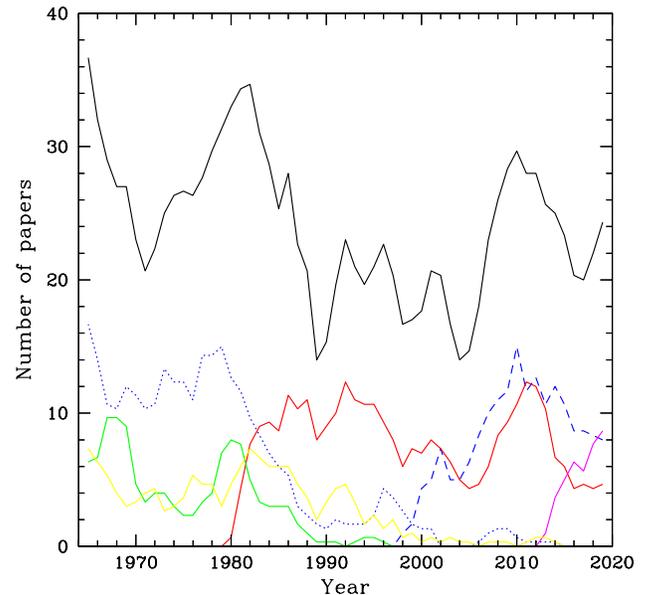}
\caption{The annual number of papers containing data from some of the
smaller telescopes of Lick Observatory between 1965 and 2019. Yearly
tallies have had a boxcar filter of width three years applied in order to partially smooth out the year-to-year fluctuations and thus make long-term
trends more apparent. Line type and color denote telescope as follows: solid black (eight telescopes combined as described in the text), blue dotted
(Crossley), blue dashed (KAIT), red (Nickel), magenta (APF), green
(Boller and Chivens 0.6 m), and yellow (Astrograph).}
\label{fg5}
\end{figure}

A breakdown of the annual number of papers that have utilized data from ``smaller'' telescopes on Mount Hamilton, i.e., other than the Shane 3 m reflector, is shown in Figure \ref{fg5}. Each curve in the figure shows a yearly tally by telescope that has been subjected to smoothing with a three-year boxcar filter. The black curve denotes papers in which were identified data derived from any of the following eight telescopes on Mount Hamilton: the Nickel 1 m reflector, the Crossley 0.9 m reflector, the KAIT, the APF, the Coud\'{e} Auxiliary Telescope, the 36 in Great Refractor, the 0.6 m Boller and Chivens reflector, and the twin Astrograph. Papers which contain data from more than one of these telescopes enter multiple times in this tally, once for each applicable telescope. Also to be noted is that some of the papers counted in this tally may also contain data from the Shane 3 m telescope, a circumstance which is particularly the case for the KAIT and Nickel telescopes. Separate counts are plotted for a number of the telescopes that contribute to the black curve in Figure \ref{fg5}. Each telescope is depicted by a combination of line type and color according to the convention: Crossley (blue dotted line),  KAIT (blue dashed line), Nickel (solid red), APF (solid magenta), Boller and Chivens 0.6 m (green), and twin Astrograph (yellow). 

Early in our 1965-2019 survey period the Crossley reflector was the
main supporting telescope at Lick Observatory, continuing a role that it
began in the 1890s. Also of significance in the first half of this period was the Boller and Chivens 0.6 m which provided a single-channel photometer as well as ITS capabilities, complementing the imaging and multi-channel scanner instruments that were run on the Crossley. Both of these telescopes began to decline in
use after the 1 m Nickel reflector was commissioned, and by the time that the
KAIT came on line just prior to the year 2000 they had ceased to be in productive use. The result is that a minimum can be seen in the productivity of the supporting telescopes at Mount Hamilton between 1990 and 2004. After 2002
the KAIT and Nickel both evince an upturn in output that peaks around 2010, a trend that can be attributed to use of the Nickel for the photometric monitoring of supernovae discovered with the KAIT.

In summary, with respect to the smaller telescopes, Figure \ref{fg5} documents how
the KAIT and APF have, in a sense, created new research niches at Lick
Observatory that have had the effect of compensating for the diminution of
the Crossley and the Boller and Chivens 0.6 m telescopes. The first KAIT papers appeared in refereed literature in 1999, with the first refereed APF paper in 2014. The rise of the KAIT, and more recently the APF, coupled with a decrease in annual papers from the Shane telescope, appear to be the reason behind a modest net decrease in the ratio between Shane and total Lick papers that may be discerned since 2005 in Figure \ref{fg3}. The Nickel has been a fairly consistent long-term presence since 1980. Although the Astrograph remained a photographic telescope up until being decommissioned, data from it continued to be used for published astrometric programs up until about 2000.

\subsection{Lick Observatory as a Test Bed for Instrumentation}

Some papers in our literature survey reflect another notable use that has been made of Lick telescopes; they have served as development test beds for new instruments that have gone to be scheduled many times for new astronomical research. The natural and laser guide star adaptive optics system on the Shane telescope \citep[e.g.,][]{Olivier1994,Max1997} is a prominent example, as
is the development of cross-dispersed echelle spectroscopy that would eventually pave the way for the HIRES spectrometer on the Keck I telescope
\citep{Soderblom1978, Vogt1981, Vogt1987}. However, other examples can also be referenced, such as the first lunar laser ranging system to collect returns from the Apollo 11 optical retro-reflector array \citep{Faller1969}, the FLITECAM infrared imager that was eventually installed on the SOFIA aircraft \citep{Smith2008}, FIRST \citep{Huby2012}, the POLISH imaging polarimeter \citep{Wiktorowicz2015}, Optical SETI \citep{Wright2001}, the VillaGEs optical adaptive optics system \citep{Gavel2008}, and the Vulcan camera \citep{Caldwell2000}. Although papers describing the development phase of such instruments have often been contained in non-refereed conference proceedings, and as such are not included in our survey (Section 3), research produced by these instruments does figure significantly in our compilation of refereed papers.

\subsection{The Plate Archive}

As with many observatories LO now maintains an electronic archive of digital
observations acquired with its telescopes. Interestingly, in the earliest years of our 1965-2019 survey period, prior to pre-electronic detectors, papers of an archival nature can nonetheless be identified. There was to an extent in these times something of a trade in antiquities, as photographic plates obtained by UC astronomers using Lick Observatory telescopes were loaned to researchers in other countries, where they were remeasured to produce new material for new publications. Such papers have been included in our compilation of Lick-based papers under the criterion that they have resulted from new measurements of Lick observational raw material. A library of archival plates dating back to the 1880s is still maintained at Mount Hamilton and occasionally has enabled new research \citep{Misch2008}.

\section{Authorship Trends}

\subsection{Trends Among First-authors}

\begin{figure}[ht]
\centering
\includegraphics[width=.50\textwidth]{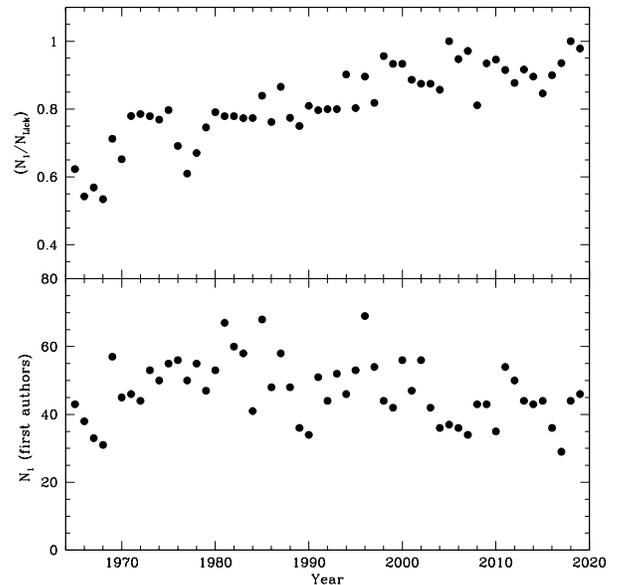}
\caption{(Lower panel) The annual variation in the number of distinct first-authors
$N_1$ in the authorship lists of Lick-based papers throughout the period 1965-2019.
(Upper panel) Ratio $(N_1/N_{\rm Lick}$) between the number of distinct first-authors on Lick-based papers to the number of Lick-based papers. The ratio has increased on average over the duration of the survey. }
\label{fg6}
\end{figure}

In most years throughout the 1965-2019 survey period multiple papers can be
identified that have the same first author. Counts were made of the number of
different first authors $N_1$ that occur for each year of the survey. This
first-author tally is shown in the lower half of Figure \ref{fg6}. The greatest numbers, amounting to almost 70 distinct first-authors, occurred between 1980 and 2000. In most years the number of distinct first authors has ranged between 30 and 60. The ratio between the number of first-authors each year and the number of Lick-based papers $(N_1/N_{\rm Lick})$ is shown in the upper half of Figure \ref{fg6}. There has been a steady increase in this ratio over the time of the survey.

Some of the greatest diversity among first authors occurred toward the end of the first peak in the LO productivity curve, although the highest value for $N_1$ seems to have been associated with the 1990's secondary peak in the publication record
(lower panel of Figure \ref{fg6}). The ratio $(N_1/N_{\rm Lick})$ has increased steadily throughout the survey period, as the number of unique first authors has shown little long-term variation while the annual number of papers has decreased on average. The pool of first authors on LO-based papers is as broad in the second half of the survey as it was in the first half.

In the 1960s and 1970s the Shane 3 m telescope in particular was a
resource of first choice for many astronomers in the University of California
system, and so it follows that such researchers would have been writing
multiple papers per year based on their Lick observations. Consequently, the
ratio of the number of first-authors to the number of papers was at its lowest early in the survey period as seen from Figure \ref{fg6}. Many UC observational astronomers reoriented their research programs to facilities such as the Keck Observatory after 1990. The UC astronomical community evidently continued to acquire observers who used Lick Observatory, either by bringing such observers directly into the UC system as graduate students, researchers, and new faculty, or by an expansion of collaborations with non-UC astronomers who are able to make use of LO data. If this second factor has been significant it would be expected to lead to a decline in the fraction of papers having a first author who is directly affiliated with a UC institution. This is discussed in Section 6.4.

\subsection{The Lick Observatory User Community}

Graduate students, faculty, postdoctoral fellows, and astronomers in research series appointments at any campus or laboratory of the University of California are eligible to apply for telescope time at Lick Observatory and to be a principal investigator (PI) on an observing proposal. Generally an ADS search of a refereed paper will provide information on the institutions with which authors are affiliated but not the type of appointments held by each author. Hence it is not possible from the ADS database alone to obtain information necessary to determine the numbers of LO-based papers that are first authored by faculty, or graduate students, etc. 

The composition of the user community of Lick Observatory can be tracked in a
more limited way from an online record of observing proposals that extends back to
2012. Over the course of 16 observing semesters extending from 2012 to 2019 
inclusive, UC faculty were the PI on 30.4\% of proposals to use the Shane telescope,
graduate students accounted for 19.5\% of Shane PIs, postdoctoral scholars for
18.5\%, research series appointees for 25.8\%, with Lick Observatory staff
making up 4.8\%. The extent to which these percentages might have been different at
earlier times we do not have firm information on. The list of co-investigators
(CoIs) on proposals for LO observing time can be extensive and not limited to UC personnel.

\subsection{Number of Authors per Paper}

\begin{figure}[ht]
\centering
\includegraphics[width=.50\textwidth]{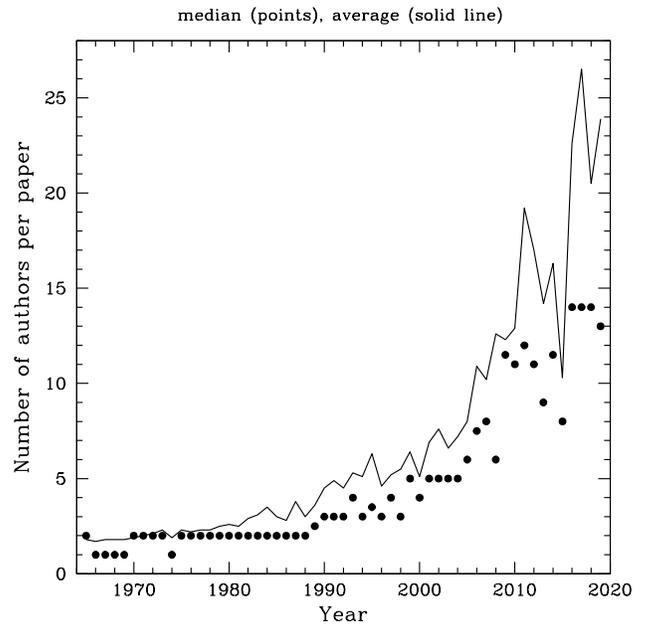}
\caption{Variation in the number of authors on Lick-related papers between 1965 and 2019. Filled points denote the median number of authors per paper for each year. The solid curve shows the average number of authors per paper.}
\label{fg7}
\end{figure}

We next turn to the total number of authors that feature on Lick-related papers. Figure \ref{fg7} shows the average number of authors per paper in a given year (solid line) as well as the median number of authors per paper per year (solid points). Both show a dramatic upturn after 1990. We suggest that there are at least several factors behind these trends. (i) Since 1990 an increase in author numbers is a common phenomenon throughout the astronomical community and is common across most journals \citep{Smith2016, Smith2017}. The user community of Lick Observatory is a partaker of this general trend. (ii) Some UC astronomers are members of large collaborations that
use either space-based or multiple ground-based telescopes, such that observations  from a LO telescope may be just one (sometimes small) component of a much larger data set presented in a group paper. Transiting exoplanet science in the Kepler era and SNe follow-up monitoring are two examples of this trend that can be discerned among LO-related papers. (iii) In the 21st century Lick Observatory has played a significant role in supporting several programs that require long-term acquisition of data, notable examples (but not the only ones) being SNe monitoring programs and active galactic nuclei (AGN) reverberation mapping programs. Movement of many UC research programs to the Keck telescopes has allowed such long-term programs to become feasible with the Shane and Nickel telescopes. Such long-term programs can also require large teams of observers, thereby reinforcing the author upturn seen in Figure \ref{fg7}. (iv) The prevalence of the internet, beginning with ftp and e-mail and more recently through wikis, messaging tools, video conferencing and collaborative cloud based editors, has allowed authors at geographically disparate universities to work collaboratively.

\subsection{Author Affiliations}

\begin{figure}[ht]
\centering
\includegraphics[width=.50\textwidth]{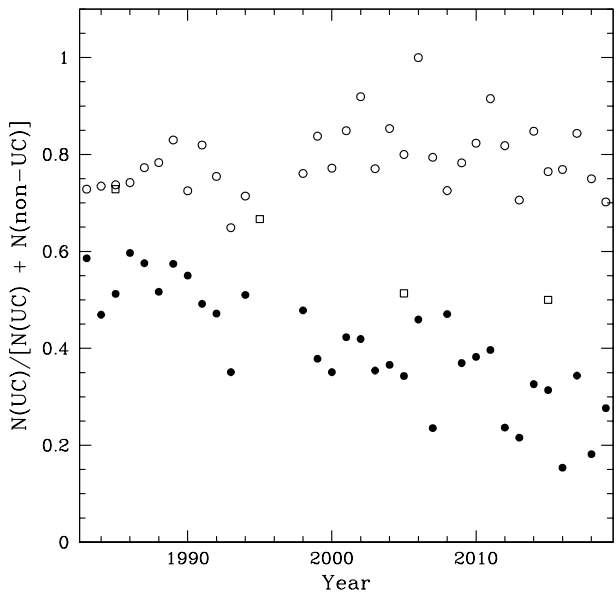}
\caption{The fraction of papers that have a UC-affiliated first author at the time of publication (filled circles), and the fraction of papers that have at least one UC-affiliated co-author (open circles) versus year since 1983. These data are based on an affiliation search made with the ADS. Open squares denote the fraction of first authors having either a UC affiliation at the time of paper publication or else a prior affiliation with a UC campus or laboratory. These points are derived from a broader range of searched material.}
\label{fg8}
\end{figure}

Using the ADS database for the library described in Section 3.2 a list of author affiliations was retrieved for most journals extending back to $\sim$ 1975. 
These listings were found to have limitations of completeness and accuracy.
Prior to 1978 the listings become very incomplete, and a gap was encountered in the data for 1995 and 1996. In a few cases papers were listed as having no UC affiliated authors when the contrary was clearly the case. Nonetheless, within the limits of
the ADS search, the affiliations list thereby obtained was used to construct two metrics: (1) the fraction of papers with first authors having a UC affiliation, and (2) the fraction of papers with any author having a UC affiliation. If an author has multiple affiliations, one of which is from UC, then the author is considered to be affiliated with the University of California. Figure \ref{fg8} shows both fractions for the period 1980-2019. Years prior to 1983, as well as 1993, 1995-1997, and 1999, have been excluded on the basis of incompleteness concerns. There is a very clear decline in the fraction of papers with UC-affiliated first authors, while the fraction of papers having at least one author with a UC affiliation has remained roughly constant. 

There is a significant complication to interpreting the trends in Figure \ref{fg8}.
Students and post-doctoral researchers may move to institutions outside the University of California system but continue to collect data from Lick Observatory using their UC mentors as a proposal lead. It is only natural for former-UC graduate students and researchers to maintain collaborative ties with their UC advisors and mentors. Our ADS library reveals many examples. Such astronomers may be publishing LO data years after they have left the UC system. Scientists who spend sabbatical periods at a UC institution, or who hold adjunct or visiting appointments for some period of time, can also fall into this category. In papers produced under such circumstances the current institution of the first author is often given, not the former UC affiliation.

There is a related effect at work in the case of graduate students, researchers and
faculty who leave the UC system for positions elsewhere. Data reduction and analysis takes some time, and causes a lag between when data are collected and when science findings are published in a refereed journal. If, for example, a researcher acquires data at Lick Observatory while working at a UC campus, but then moves to a new position at a non-UC school, the final affiliation listed on a LO-derived paper may reflect the more recent non-UC school. Again, our ADS library reveals many instances of this, and it can lead to a LO-based paper on which none of the authors list a UC affiliation. Figure \ref{fg8} shows that up to 20\% of papers may fall into the no-UC-affiliations category in a given year. The use of archival LO data sets by non-UC researchers can also account for some fraction of no-UC-affiliated papers. For example, some papers which have used on-line data from the UCB KAIT supernovae survey program fall into this category. 

The declining fraction of UC first authors indicates that non-UC co-investigators 
on observing proposals to LO may sometimes be driving the proposed research. However, for the above reasons, some of those non-UC co-investigators may have former close connections to the UC system. Either the fraction of such CoIs is increasing, or the role of non-UC CoIs without a former UC affiliation is increasing. We have tried to test these two possibilities by making a more intensive search for former-UC affiliations among first authors for the years 1985, 1995, 2005, and 2015. Material that has been drawn upon includes the ADS PhD thesis database (to identify former graduate students), personal websites, online curricula vitae, and a list of former alumni posted online by the UCLA Division of Astronomy and Astrophysics. Using such material we have endeavored to calculate for the above four years the fraction of first authors having either a current (at the time of paper publication) or prior affiliation with a UC campus or laboratory. These fractions are shown by open squares in Figure \ref{fg8}. 

The squared points in Figure \ref{fg8} are offset above the filled circles, thereby
demonstrating that indeed there has been a consistent population of non-UC first
authors among the clientele of LO data users who nonetheless have formerly been affiliated with a UC institution. However, the additional fact that these points also evince a decreasing trend with time indicates that the fraction of first authors who have never held a UC institutional affiliation has also been increasing.
This is presumably a reflection of collaborative trends among UC astronomers, such as those discussed in Sections 6.1 and 6.3.

\section{Demographics by Research Area}

\subsection{Galactic and Extragalactic Astronomy at Lick Observatory}

With the intent of gaining a very broad-brush overview of the astronomical
research facilitated at Lick Observatory, an attempt was made to classify the more than 3200 papers in our survey between 1965 and 2019 into the two generic areas of Galactic (a paper was focused on objects within the Milky Way galaxy) and extragalactic (the objects investigated are external to the Milky Way) astronomy. Admittedly this is a very coarse classification, however, it does permit some trends to be discerned.

\begin{figure}[ht]
\centering
\includegraphics[width=.50\textwidth]{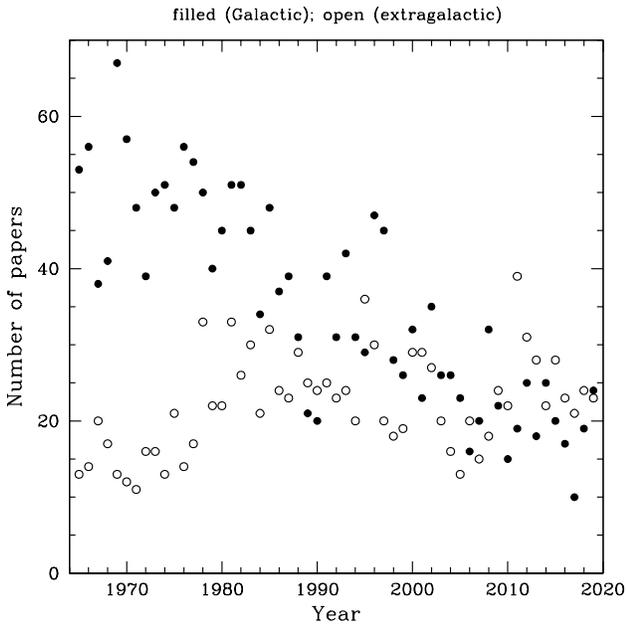}
\caption{The annual number of papers derived from Lick Observatory
telescopes between 1965 and 2019 categorized according to whether the
astronomical objects reported upon were within the
Galaxy (filled circles) or were extragalactic (open circles).}
\label{fg9}
\end{figure}

The number of Galactic and extragalactic papers per year derived from LO
observations is plotted in Figure \ref{fg9}.\footnote{Some papers were classified as being instrumental in nature, and as such the sum of the two categories in Figure \ref{fg9} may be less than the number $N_{\rm Lick}$ in a given year.} Between 1976
and 1980 there was a noticeable upturn in the number of extragalactic papers,
which almost doubled from $\sim 15 \pm 5$ per year at the start of the survey
period to $\sim 25$-30 between 1980 and 1986. Over this time there is no great
systematic change in the output of Galactic papers although these do show much
scatter between 40 and 60 per year. The greatest number of Galactic papers
occurred in 1969, which clearly stands out as a maximum for such publications
over the entire 55 yr survey. The start of the era of the ``first peak'' thus coincides with a maximum in Galactic papers, while the annual number of extragalactic
papers increased throughout this era. Thus it is possible that the start of
the first-peak period was driven by stellar astronomy at Lick Observatory, but was
sustained into the 1980s by the rise of extragalactic observations. The upturn in extragalactic papers at this time was facilitated by the installation of the cassegrain spectrograph and image tube scanner on the Shane telescope.

Since 1980 the typical number of extragalactic papers per year has fluctuated
between $\sim 20$ and 35 with no obvious systematic increase or decrease. However, the annual number of Galactic papers has fallen systematically since the end of the first peak in the LO publication record. The increasing emphasis on extragalactic papers from Lick Observatory data has lead to a systematic decrease in the relative number of Galactic to extragalactic papers per year. There is a striking drop in this ratio between 1965 and 1980, after which the decline in the fraction of Galactic papers became more gradual. Between 1990 and 2010 the ratio between the number of Galactic and extragalactic LO-associated papers was close to unity. Over this twenty year period exoplanet studies became a highly visible part of astronomy with the Shane telescope. Since 2010 the annual number of extragalactic LO papers has more often than not exceeded that of Galactic papers.

\begin{figure}[ht]
\centering
\includegraphics[width=.50\textwidth]{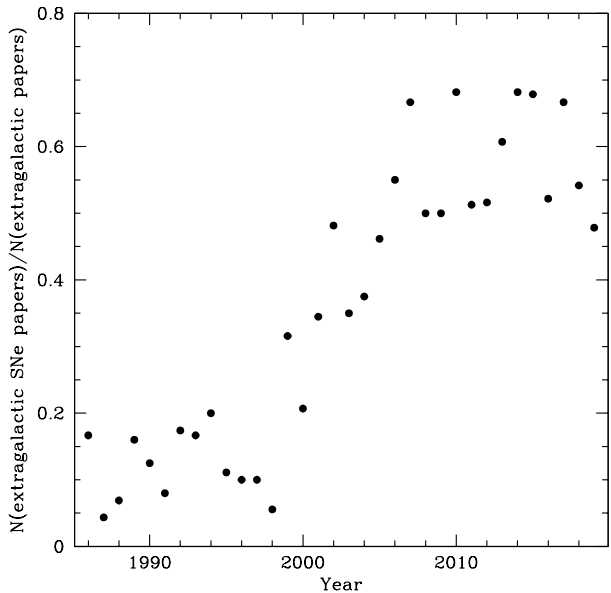}
\caption{Ratio between the number of extragalactic supernova papers and all
extragalactic papers derived each year since 1985 from telescopes at Lick
Observatory. }
\label{fg10}
\end{figure}

There is, however, a significant distinction to be made between the types of
extragalactic objects studied at Lick Observatory within the first and second halves of the survey period. From 1965 though the 1980s there was considerable spectroscopic and photometric work done on ``normal'' galaxies. In addition, the observation of active galactic nuclei and quasars was prominent within the University of California system. By contrast, since the year 2000 a large fraction of extragalactic papers arising from observations at LO have centered upon extragalactic supernovae rather than the intrinsic properties of galaxies themselves. Figure \ref{fg10} documents this trend. An effort was made to separately count for each year the number of LO-based papers that were comprised of observational studies of {\it extragalactic} supernovae. Plotted in Figure \ref{fg10} is the annual ratio between the number of extragalactic SNe papers and the number of all extragalactic LO papers. Prior to 1985 the number of such SNe papers per year was small and so is not shown in the figure. Since 1986 the fraction of all extragalactic papers that are devoted to extragalactic SNe has consistently increased, and in some years percentages near 70\% have been reached.

Groups using LO for SNe research have been active at UC Berkeley (UCB), including one
very prolific group that has been associated with Alex Filippenko. The Lick Observatory Supernova Search (LOSS) program by this group \citep{Filippenko2001,Li2011} saw the placement of the KAIT telescope on Mount Hamilton, from which has followed a very productive enterprise of extragalactic SNe discovery. There are also other active SNe programs within the UC system, with current groups at the Lawrence Berkeley National Laboratory, and in astronomy departments at UC Santa Cruz, UC Davis, and UC Santa Barbara \citep[e.g.,][]{deJaeger2019, Hosseinzadeh2018,Kilpatrick2018,Nugent2011, Shivvers2019}. Follow-up spectroscopy and photometry of SNe has been for some
years a significant consumer of observing time on the Shane and Nickel telescopes.

Another high-profile research area for Lick Observatory telescopes has been
the study of exoplanets. This was originally begun through the radial velocity
monitoring program initiated by \citet{Marcy1993, Marcy1996} with the Hamilton high-resolution spectrometer \citep{Vogt1987}. In recent years exoplanet radial velocity studies at LO have migrated to the Automated Planet Finder telescope \citep{Vogt2014}. Nonetheless, observational programs such as adaptive optics imaging of exoplanet host stars and young proto-stellar disks have ensured that exoplanet-related science has remained on the Shane observing schedule. Between 1996 and 2013 there were typically from 2 to 8 exoplanet papers produced annually from LO observations, with a peak of 12 in 2014. Nonetheless, each year since 2001 LO exoplanet papers have been fewer in number than extragalactic SNe papers.

\subsection{Author Networks and the Second and Third Peaks in the Publication Record}

Documentation of a more in-depth nature of both the community using Lick Observatory telescopes, and the types of astronomical objects studied, can be had by applying the ``Author Network'' tool within the SAO/NASA ADS system to the library of Lick-based papers described in Section 3.2. The Author Network function searches a specified set of papers and returns an author list that is conveniently grouped according to the identification of collaborative networks. Reasonably well-defined astronomical subjects can be associated with some, but not all, of the identified networks. Among $\approx 1000$ Lick-based papers covering a period from 2001 to 2018, the most numerous types of subject matter identified, listed in order of prevalence, were concluded to be: (i) extragalactic supernovae, (ii) extra-solar planets, (iii) a heterogeneous combination of adaptive optics observations and instrumentation, young stars, and stellar disks, (iv) quasars and active galaxies including reverberation
monitoring of AGN, and (v) low-mass stars and brown dwarfs. There are many more papers, however, which are addressed to objects not covered by any of these five groups. Author networks can be of assistance in helping to identify the causes behind the peaks in the LO productivity curve.

\begin{figure*}[ht]
\centering
\includegraphics[angle=270,width=1.05\textwidth]{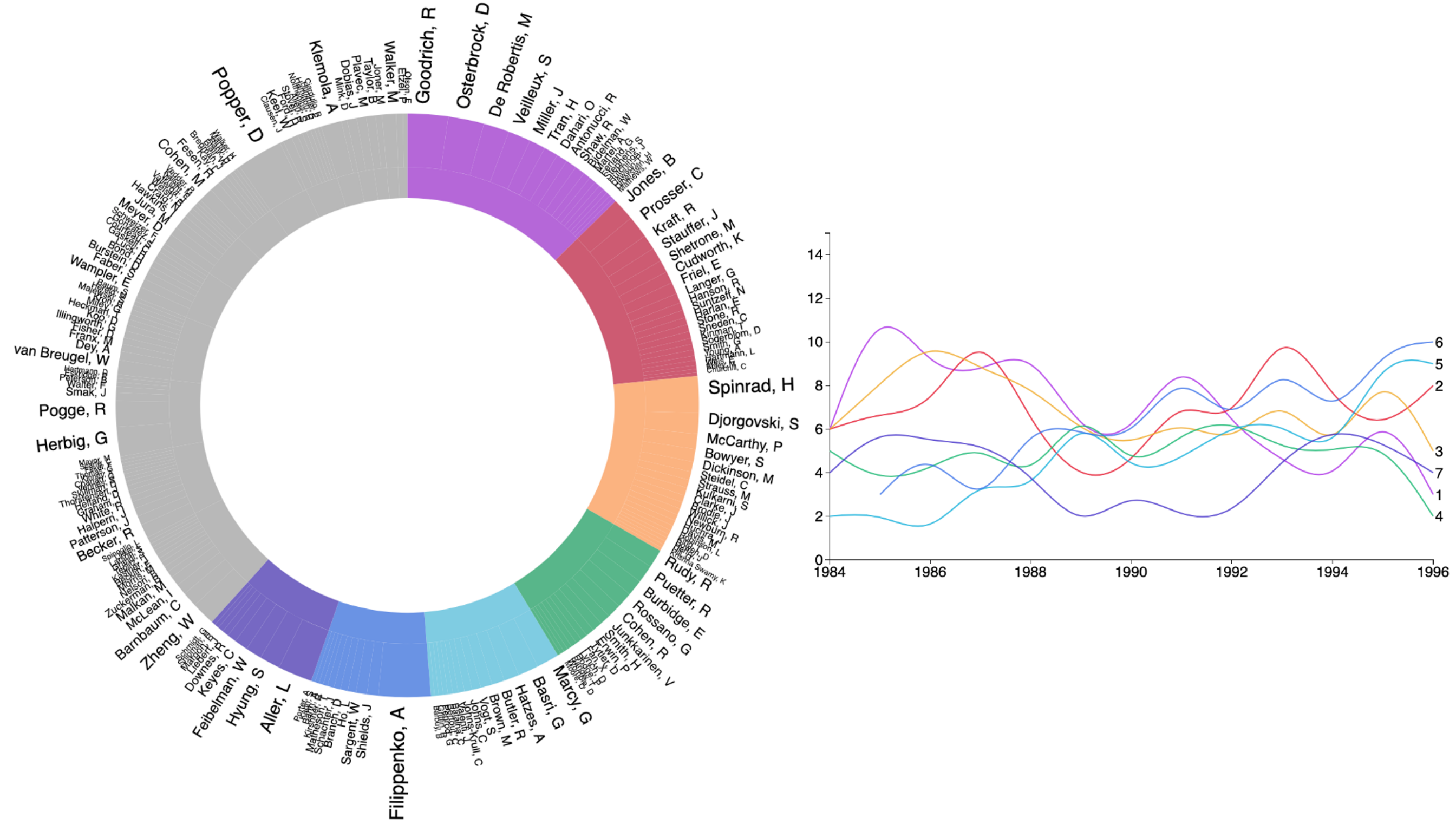}
\caption{Author network for years 1984-1996, leading up to the second peak seen in Figure \ref{fg1}. This figure was produced by use of ADS tools. The chart in the left panel depicts the author networks identified by the ADS software. In the right panel the number of papers per year from network groups 1 through 7 is plotted.}
\label{fg11}
\end{figure*}

\begin{figure*}[ht]
\centering
\includegraphics[angle=270,width=1.05\textwidth]{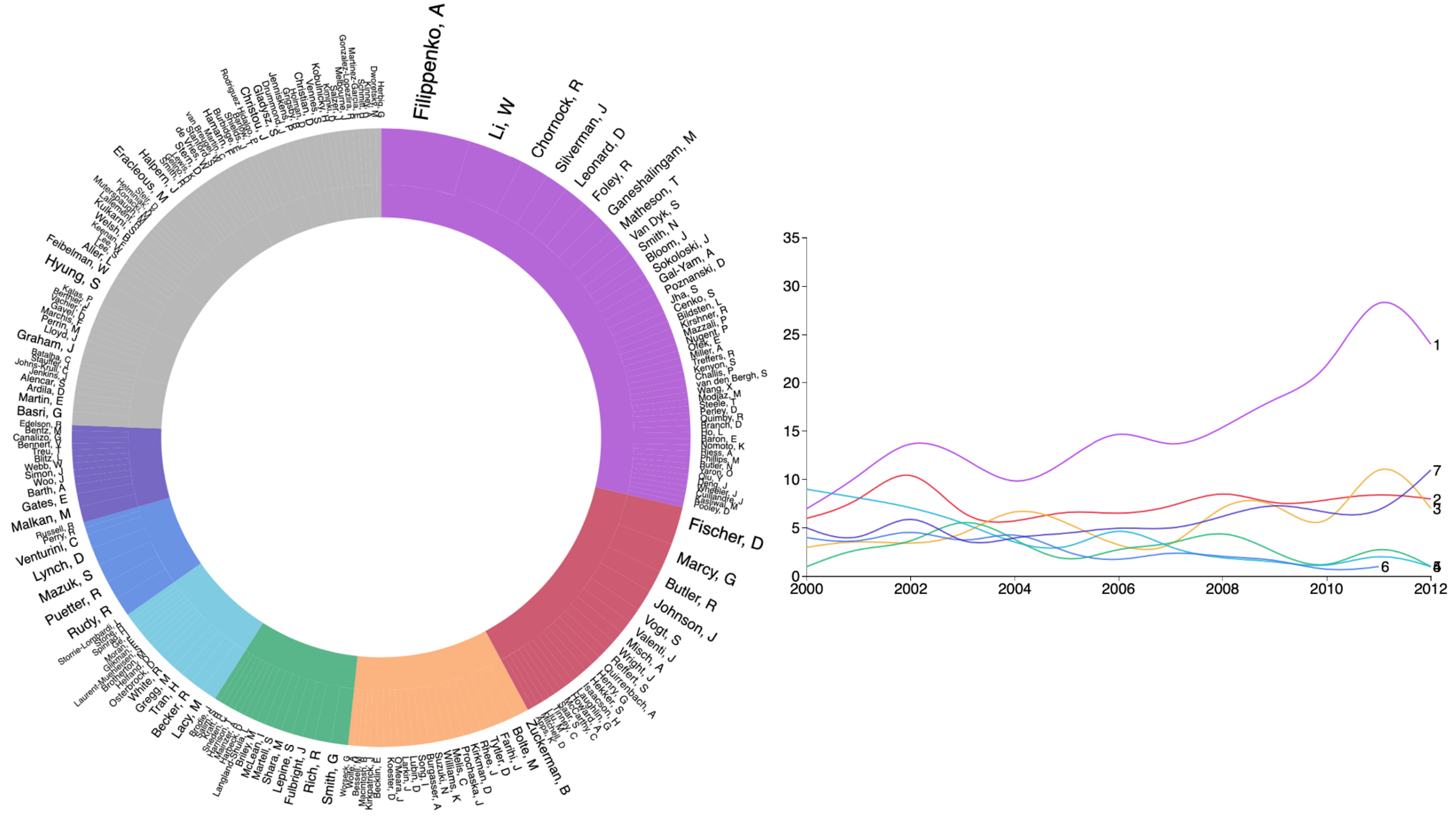}
\caption{Author network for years 2000-2012, leading up to the third peak seen in Figure \ref{fg1}. The figure was produced by use of ADS tools. Left panel: author networks identified by the ADS software. Right panel: number of papers per year from network groups 1 through 7.}
\label{fg12}
\end{figure*}

An author network is presented in Figure \ref{fg11} for the period 1984-1996, which leads up to and contains the second productivity peaks seen in Figures \ref{fg1} and \ref{fg2}. Between 1990 and 1996 the annual output of papers from groups 6, 5, and 2 identified in Figure \ref{fg11} increase on average to be the largest contributors around the time of the 1996 second peak. Group 6 identified in the figure, combined with group 1, indicate a substantial number of astronomers, particularly at UCB and UCSC as well as UCSB, who were engaged in the spectroscopy of active galaxies, including Seyfert and weak-lined AGN (the so-called LINER galaxies). The rise of group 6 at UCB can be noted during this period, whereas that of group 1 decreases through the period. Their work was based on low-resolution cassegrain CCD spectroscopy at the Shane telescope. Groups 2 and 5 in Figure \ref{fg11} both comprise high-resolution spectroscopic studies made with the Hamilton echelle spectrometer at the Shane telescope. These groups emphasize stellar abundances in the first case, with a notable center around UCSC, and stellar activity and young stars in the second case, centered around UCB. Thus it seems that the second peak in the Shane productivity curve of Figure \ref{fg2} was associated with a rise in papers based on CCD spectroscopy both at low and high resolution,
driven by groups 6, 5 and 2 in the author network. This supports the discussion of Section 5.2. As with the first peak, the second peak is not so much defined by a distinctive scientific niche, but with patterns of instrumentation use.\footnote{As regards other groups in Figure \ref{fg11}, low-resolution extragalactic astronomy, sometimes of a higher-redshift nature, attends group 3. Group 4 represents the work of southern UC campuses, in particular UCSD, on the subject of quasar research. As with groups 1 and 6, these two groups were active users of the UV-Schmidt and the successor Kast low-resolution spectrographs at the cassegrain focus of the Shane. Group 7 in Figure \ref{fg11} prominently features the work of Aller and Hyung at UCLA, whose work on planetary nebula abundances benefited from both the high- and low-resolution spectroscopic capabilities of the Shane telescope.} 

A second author network shown in Figure \ref{fg12} covers the years from 2000 to 2012 and leads up to the third, but lesser, peak in the Lick productivity curve that occurs around 2012. The right-hand panel of Figure \ref{fg12} provides a compelling explanation for the origin of this third peak. It is due to a rise in extragalactic supernova astronomy within the UC community that has been prominently centered around UC Berkeley. Of the seven author groupings in Figure \ref{fg12}, it is group 1, which can be clearly identified with observations of extragalactic supernovae, that not only produced the greater number of papers per year, but also sustained an increasing number of papers per year throughout the 2000-2012 period. This is an era of KAIT supernovae discoveries and their follow-up at the Shane and Nickel telescopes. Figure \ref{fg10} provides further verification of the rise of this specific research field at Lick Observatory. Follow-up spectroscopy of SNe is mainly
accomplished with the Kast spectrograph, and the upturn in Kast usage at the time of
the third peak (Figure \ref{fg4}) can be attributed to this effect.
The other major group in Figure \ref{fg12} (Group 2), which  contains the exoplanet community within UC, has a rate of production of LO-based papers that held reasonably constant throughout the plotted time period.\footnote{Groups 5 and 7 within Figure \ref{fg12} depict the continuation of active galaxy programs at Lick Observatory into the 21st century. Infrared astronomy at LO is not isolated to a single group in Figure \ref{fg12} but is spread through groups 3, 4 and 6, and the gray segment.} 

\subsection{Attempt at a Finer Classification of Papers by Research Subject}

In an effort to facilitate a more detailed look at the science topics covered by LO-data papers we experimented with devising a finer grid than a mere Galactic/extragalactic distinction for classifying papers by subject matter. Our approach has been guided by the style of journal key words. However, since this concept post-dates the 1965 start of our literature survey an attempt was made to formulate a key word style of classification that accommodates many of the themes conspicuous in the LO-based literature. Since categorizing papers in this way is rather laborious and occasionally ambiguous we limited our efforts to doing annual classifications at five-year intervals. The results are presented in Table \ref{table3}, with footnotes for several of the categories, and they illustrate a number of the trends
discussed above. The decline in stellar astrophysics with LO telescopes is well documented in Table \ref{table3}, as is the rise of extragalactic SNe and exoplanet studies. Active galaxies have been a fixture throughout the 1965-2019 period, particularly in the middle years. Both Galactic interstellar medium studies and investigations into fundamental properties of external galaxies once held prominent roles on the observing calendar of Lick Observatory but both have declined notably. 

\section{Conclusions}

Upon the above considerations of the publication record of Lick Observatory it is suggested here that the scientific productivity of the observatory has been strongly
influenced by at least three factors: (i) telescopes, (ii) instrumentation, and (iii) a user community that has chosen to place particular emphasis on specific scientific areas. These three factors have interacted with each other.

\subsection{Telescopes}

Having a large telescope, the Shane 3 m reflector, which was the second-largest
optical telescope on Earth during the earliest years of our 1965-2019 survey
period, has clearly been a major factor in setting the productivity of Lick
Observatory. The Shane has remained the largest telescope on Mount Hamilton
throughout the succeeding eras of larger 4 m class and then 8-10 m class
optical telescopes worldwide, and it has dominated the research
output from Lick Observatory throughout the 55 yr survey period.

Nonetheless an ecosystem of smaller telescopes has consistently accounted
for 20-40\% of the research output from Lick Observatory. In the first
half of the survey period the general purpose 0.9 m Crossley and 0.6 m
Boller and Chivens reflectors supported a wide range of observing programs.
The Double Astrograph, by contrast, was a special-purpose telescope devoted
to a niche area of astronomy, namely a large stellar astrometry program.
In the second half of the survey period, the 1.0 m Nickel became a natural
general-purpose successor to the Crossley, however, major contributions have
been made by two specific-purpose telescopes designed to perform niche
science, namely the KAIT and the APF. Thus the publication record of
Lick Observatory provides an illustration of how, in the era of spaced-based
and large ground-based optical telescopes, smaller telescopes can remain
competitive by being devoted to well-defined scientific objectives.
\subsection{Instrumentation}

The publication record during 1970-2000 for Lick Observatory shows that
having a large telescope is just one component to supporting a diverse
scientific community. The record also shows the impact of introducing new
technology into the instruments used on the Lick telescopes. The clearest example of this in our 55 yr survey is the introduction of the image-tube, image-dissector scanner and its combination with an optimized cassegrain spectrograph \citep{Miller1980} on the Shane telescope. This instrument combination drove the first and largest productivity peak seen in the Lick publication record. Around the time of the second peak the highest publication rate from the Hamilton echelle spectrometer occurred, the annual number of papers derived from the newly-introduced Kast spectrograph rose sharply, and the first papers came from the UCLA Gemini camera. In the case of the Shane telescope such upgrades helped offset factors such as the world-wide deployment of larger 4 m class telescopes and the increasing brightness of the Mount Hamilton night sky.

Although the first publication peak for Lick Observatory was fairly broad, on the order of a decade, the second peak was more limited. Perhaps in part the second peak was truncated by UC astronomers redirecting their research programs to take advantage of access to the much larger Keck I telescope. Since circa 1980 the average productivity of Lick Observatory has shown a progressive decline upon which has been superposed the second and third peaks in the publication record. If one might draw a lesson from the second peak it may be that instrument development can only go so far
in keeping a telescope, of what is now a moderate aperture, competitive, once a user community gains access to an enhanced range of observational resources. By contrast, the third peak in the Lick Observatory publication record seems to have been 
driven by a specific niche science and the introduction of a telescope to support that niche.

\subsection{Specialized Science Directions}

Several research emphases can be identified in the Lick publication record (Table \ref{table3}). The most enduring long-term sub-discipline supported by Mount Hamilton telescopes has been the study of nuclear activity in galaxies, ranging from QSOs, to radio galaxies, Seyferts, and lower-activity systems. Active galaxy astronomy is a subject that has been pursued across almost all campuses of the University of California system.

Despite the variety seen in Table \ref{table3}, the specific science that lead to
the third peak in the LO publication record is the study of
extragalactic supernova. It is an instructive example of niche science for
at least three reasons. Firstly, it has been facilitated by the placement of
a low-cost special-purpose survey telescope on Mount Hamilton for SNe
discovery. This is the first example of LO becoming what might be termed a
``cliente'' observatory, a trend that a number of major observatories around
the world have partaken of, often to a much greater extent than Mount
Hamilton. Second, a large fraction of the subsequent science has come
from follow-up photometry and spectroscopy that has been carried out with
general-purpose instrumentation already in place on LO telescopes, most
notably the Kast spectrograph on the Shane and a CCD imaging camera on the
Nickel telescope. The third and decisive factor that initiated and has driven
extragalactic SNe niche science at Lick Observatory has been a stable
long-term group centered at the University of California at Berkeley under
the leadership of Alex Filippenko, although other supernova groups within the
UC system have also employed the Shane telescope.

It may be that specialized science directions are a key to the further longevity of Lick Observatory. The SNe example indicates that niche science can come about via
at least two approaches. One is the installation of new special-purpose
telescopes committed to a niche science, the APF is the most recent example of
this at Mount Hamilton. A second approach is the commitment of large blocks of
observing time on existing telescopes, the APF illustrates this for exoplanet
science, while the Shane telescope has been used in this mode for SNe follow-up observations and AGN reverberation monitoring. With the era of the LSST approaching it might be that other forms of transient follow-up will present opportunities at Mount Hamilton. A pressing question in planning for the future of Lick Observatory will be to decide whether to opt for predominantly niche science, or whether to maintain some significant capability for short-term observing programs having a diverse range of astronomical themes, as has been the traditional mode of telescope time allocation in the past.   These options can have significant impact on the scheduling model a telescope operates under, queue vs. classically scheduled, and on the levels of technical support needed to operate the facility, e.g. instrument changes and availability for visiting instrumentation.

%\ack
\begin{acknowledgements}
We wish to recognize the technical and support staff of Lick Observatory,
many of whom reside on Mount Hamilton. It is a pleasure and privilege to 
be associated with such a dedicated and talented group of people. 

At the time this paper was being written Lick Observatory survived a major fire event
that had seriously threatened the infrastructure of the observatory. We are
deeply grateful to the efforts of CalFire personnel.

This research has made use of the SAO/NASA Astrophysics Data System. The ADS is operated by the Smithsonian Astrophysical Observatory under NASA Cooperative Agreement NNX16AC86A. We thank Steve Allen for supplying data from the digital archive of Lick observing proposals.
\end{acknowledgements}

\bibliographystyle{apj}
\bibliography{ms}

\newpage

\begin{table}[ht]
  \centering
  \caption{Telescopes in Use at Lick Observatory 1965-2019}
  \label{table1}
  \begin{tabular}{ccc}
\hline\hline
Telescope  &  Commissioned  &   Status in 2019   \\
\hline
3.0 m Shane      &  1959    &      in service   \\
2.3 m APF        &  2013    &      in service    \\
1.0 m Nickel     &  1979    &      in service    \\
0.9 m Crossley   &  1896    &       most recent science obs 2009   \\
KAIT             &  1996    &      in service    \\ 
Coud\'{e} Auxiliary Tel     &  1972    & most recent science obs 2014      \\
36 in Refractor  &  1888    &       most recent science obs 2012 \\ 
0.6 m B\&C       & c~1961   &     dismantled 1995   \\
0.5 m Tauchmann  & c~1955   &       inactive      \\
20 in Astrograph &  1941    &   most recent science obs 2000 \\
12 in Clarke     &  1881    &     removed 1979   \\
\hline\hline
  \end{tabular}
\end{table}

\begin{table}[ht]
    \centering
    \caption{Instruments on the Shane Telescope 1965-2019}
    \label{table2}
    \begin{tabular}{cccc}
\hline\hline
Focus      &  Instrument   &  Main dates of Productivity   &   Notes  \\
\hline
cassegrain & Kast spgh          &  1992-2019     &  \\
     ...   & pre-Kast CCD spgh  &  1984-2002     &  Ref: \cite{Miller1980} \\ 
     ...   & ITS + spghs        &  1972-1998     &  first installed 1970 \\
     ...   & adaptive optics    &  1997-2019     &  Ref: \cite{Max1997} \\
     ...   & UCLA Gemini IR camera &  1995-2013  &  Ref: \cite{McLean1994} \\
coud\'{e}  & spectrometer       &  1965 - c1993  &       \\
           & Hamilton CCD echelle  &  1987-2019  &  Ref: \cite{Vogt1987}   \\
prime      & PF spectrograph    &  1965-1975     &     \\
           & plate camera       &  1965-1986     &     \\
           & Wampler scanner    &  1966-1979     &      \\
           & pe photometer      &  1965-1976     &      \\
           & PF CCD camera      &  2000-2012     &    \\
\hline\hline          
    \end{tabular}
\end{table}

\newpage

\begin{table}[ht]
   \centering
   \caption{Representation of Research Areas in LO-Based Papers}
   \label{table3}
   \begin{tabular}{lrrrrrrr}
   \hline\hline
           &  Years  &  Years  &  Years  &  Years  &  Years  &  Years  &  Years  \\
           &  1965  &   1969   &  1975   &  1985   &  1995   &  2005   &  2015   \\
           & +1966   & +1970   & +1979   & +1990   & +2000   & +2010   & +2017   \\
           &         & +1971   & +1980   &         &         &         & +2018   \\
           &         &         &         &         &         &         & +2019   \\
   \hline 
 Category  &   \%    &   \%    &   \%    &    \%  &   \%    &   \%    &    \%   \\     
   \hline       
Instrumentation:                        &  2.9 &  0.5 &  1.0 &  0.8 &  1.6 &  1.4 &  2.9 \\
Solar System:                           &  7.2 &  0.5 &  3.0 &  5.7 &  3.2 &  6.8 &  4.6 \\
Exoplanets, SETI:                       &  0.0 &  0.0 &  0.0 &  0.0 &  6.3 & 10.8 & 18.4 \\
Stars: clusters, astrometry, motions    &  4.3 &  7.5 &  4.5 &  6.5 &  2.4 &  1.4 &  0.0 \\
Stars: binary, multiple                 &  9.4 &  8.0 &  4.0 &  6.5 &  6.3 & 10.8 &  5.7 \\
Stars: variable$^1$                     & 15.2 & 11.9 & 17.2 &  4.9 &  3.2 &  6.8 &  1.7 \\
Stars: abundances, chemically peculiar$^2$    & 13.0 & 17.9 &  6.6 &  5.7 &  3.2 &  1.4 &  4.0 \\
Stars: young stars, activity, winds$^3$ &  2.9 &  1.5 &  9.6 &  7.3 &  5.6 &  2.7 &  2.9 \\
Stars: other$^4$                        & 14.5 & 19.4 & 10.6 &  3.3 &  7.1 &  8.1 &  0.6 \\
Galactic ISM$^5$: HII regions, PNe      & 11.6 & 12.9 &  9.1 & 12.2 &  7.9 &  2.7 &  0.6 \\
SNe$^6$, SNRs$^6$, novae$^6$:           &  2.2 &  1.5 &  1.5 &  4.9 & 11.1 & 31.1 & 37.4 \\
Galaxies: normal, individual            &  7.2 & 11.9 &  9.1 &  6.5 &  4.0 &  4.1 &  1.7 \\
Galaxies: active, AGN, QSOs, radio      &  8.7 &  6.5 & 20.2 & 30.1 & 27.0 &  9.5 & 17.2 \\ 
Galaxies: other$^7$                     &  0.7 &  0.0 &  3.5 &  5.7 & 11.1 &  2.7 &  2.3 \\
   \hline\hline
   \multicolumn{8}{l}{$^{1}$includes Galactic pulsars and stellar x-ray sources but not novae}\\
   \multicolumn{8}{l}{$^{2}$includes properties of carbon stars}\\
   \multicolumn{8}{l}{$^{3}$includes mass loss and some forms of circumstellar shells but not PNe + young star disks}\\
    \multicolumn{8}{l}{$^{4}$includes general photometry, spectroscopy, spectral types, effective temperatures}\\
    \multicolumn{8}{l}{$^{5}$ includes diffuse interstellar bands, temperature of ISM}\\
    \multicolumn{8}{l}{$^{6}$includes both Galactic and extragalactic supernovae (SNe), supernova remnants (SNRs), and novae}\\
    \multicolumn{8}{l}{$^{7}$includes motions, redshifts, clusters, circumgalactic medium, absorption line systems}\\
   \end{tabular}
\end{table}

\end{document}